# Engineered Dual BIC Resonances in Hybrid Metasurfaces for Controlled Photoluminescence Amplification

Omar A. M. Abdelraouf*, Mengfei Wu, Hong Liu*

**ABSTRACT**

The development of miniaturized light sources with tunable functionality is crucial for advancing integrated photonic devices, enabling applications in quantum computing, communications, and sensing. Achieving tunable light emission after device fabrication remains a significant challenge, particularly when efficient amplification is required. Hybrid metasurfaces, which integrate several nanostructured materials to form optical resonators, have emerged as promising candidates to overcome these limitations by providing a high degree of flexibility in emission control and enhanced amplification. In this work, we demonstrate tunable amplified photoluminescence (PL) in nanocrystalline silicon (nc-Si) quantum dots (QDs) embedded in a hybrid metasurface consisting of amorphous silicon (a-Si) and a low-loss phase change material (PCM) antimony trisulfide ($Sb_2S_3$). The nc-Si QDs maintain a high PL efficiency and stability at elevated temperatures, offering reliable and tunable phase transitions in the PCM. The hybrid metasurface supports dual quasi-bound states in the continuum (BICs) to achieve $Q$-factors up to 225. The dual BIC cavity enables tunable amplified PL by a factor of 15 with a wavelength shift of up to 105 nm via dimensional modulation. Meanwhile, all-optical tunable PL emission across a 24 nm wavelength range has been achieved when PCMs are tuned from the amorphous to crystalline phase. Furthermore, we propose a high $Q$-factor metalens to focus the tunable amplified PL, extending the diffraction-limited focusing tunability into the near infrared (NIR). This work paves the way for highly efficient quantum light sources using reconfigurable nanophotonic devices in next-generation photonic systems.

The integration of nanocrystalline silicon (nc-Si) quantum dots (QDs) into optoelectronic devices has gained considerable attention due to its ability to overcome the intrinsic limitations of bulk silicon as a light-emitting material.[1-3] Unlike bulk silicon, which suffers from an indirect bandgap and is dominated by non-radiative recombination, nc-Si QDs is a direct bandgap material with quantum light confinement.[4-6] That enables efficient stimulated light emission, high photoluminescence (PL) efficiency, and exceptional chemical and thermal stability.[7-9] Despite these advantages, tuning and amplifying the PL emission of nc-Si QDs in real time in a device remains a significant challenge.[10-12] Once a device is fabricated, the fixed geometry and inherent material properties limit the ability to dynamically tune the light emission, restricting its use in applications requiring on-demand functionality. This challenge highlights an urgent need for solutions capable of dynamically engineering and amplifying light emission after fabrication, particularly in the visible and near infrared (NIR) spectral ranges.[13-15]

Nanophotonics offers a promising approach to enhancing and controlling light emission in fluorescent materials by employing resonant nanostructures.[16-18] These structures can support modes with extremely high-quality factors ($Q$-factors), such as bound states in the continuum (BICs), thus enabling enhanced light-matter interactions.[19-21] BIC resonances allow for strong localization of electromagnetic fields, thereby enhancing the rate of spontaneous emission and significantly amplify PL intensity. However, to fully harness the potential of nc-Si QDs and meet the growing demands for advanced photonic devices, the development of tunable systems is essential. This necessitates the incorporation of materials such as phase-change materials (PCMs) that can dynamically adjust their optical properties inside nanophotonic devices.[22-24] By integrating tunable PCMs with resonant nanostructures, it becomes possible to dynamically control PL enhancement and emission characteristics for fabricated devices, offering a new level of adaptability.

Phase-change materials, such as antimony trisulfide ($Sb_2S_3$), offer distinct advantages in tunable photonic systems. $Sb_2S_3$ exhibits high refractive index ($n > 4$) and low optical losses ($k < 0.01$) in the visible spectrum, making it a competitive material for high efficiency light-emitting applications. In addition, it provides ultrafast optical switching capabilities for high-speed optoelectronic devices.[25-27] However, $Sb_2S_3$ undergoes phase transitions at relatively high temperatures compared to other PCMs and semiconductor nanocrystals. This poses challenges for the stability of the emitting materials during multiple switching cycles. In contrast, nc-Si QDs demonstrate robust thermal stability, withstanding temperatures up to 800°C.[28] This high-temperature resilience makes nc-Si QDs an ideal candidate for integration with $Sb_2S_3$ in hybrid systems, providing an adequate stability during phase transitions while simultaneously maintaining efficient light emission.

Hybrid metasurfaces, consist of more than material, offer two key functionalities for achieving tunable amplified PL emission.[29-31] First, they act as resonant cavities that enhance light-matter interactions and amplify PL through BIC resonances. Second, the $Sb_2S_3$ film functions as an active modulator which can tune the amplified PL wavelength emission upon phase transition. This hybrid approach enables the dynamic control over the emission properties to effectively addresses the challenges of device tunability. The realization of cavity-enhanced light-matter interactions and optical tunability through low loss PCMs offers a versatile and reconfigurable platform enabling precise and active control over light emission.[32-34]

In this study, we demonstrate a dual BIC resonance-enabled hybrid metasurface which achieve tunable amplified PL (TAP) with a wavelength shift of up to 105 nm through sweeping device geometries. By leveraging a hybrid nc-Si/amorphous silicon (a-Si)/$Sb_2S_3$ cavity, we overcome the limitations of current light-emitting nanophotonic devices. Moreover, we achieve an amplified PL of up to 15 times with a tunable emission wavelength spanning 24 nm through the phase change of $Sb_2S_3$. Additionally, the introduction of a high-Q-factor metalens enables focusing of the enhanced emission, resulting in a broadband focused tunable light source in the NIR spectrum. The development of this hybrid approach demonstrates its potential to overcome key challenges related to efficient, tunable light emission and material stability, opening new avenues for applications in quantum technologies,[35] optical communications,[36] and advanced sensing.[37] Through meticulous engineering of the hybrid structures and careful optimization of the resonant conditions, we present a robust, scalable solution for next generation nanophotonic devices.

**RESULTS AND DISCUSSION**

Figure 1 shows the concept of TAP from metasurfaces with dual BIC resonances for compact tunable light sources, and the 3D schematic of the proposed hybrid metasurface is illustrated in Fig. 1a. Each meta-atom (unit cell) consists of two elliptical pillars, and each pillar is made of a stack of three materials, including the bottom a-Si ellipse of thickness of 400 nm, top $Sb_2S_3$ ellipse of thickness of 130 nm, and in between a buffer layer of hydrogen silsesquioxane (HSQ) of thickness of 50 nm. The entire metasurface is coated with a layer of $Al_2O_3$ of thickness of 10 nm. The incident continuous-wave (CW) laser has a horizontal polarization along the $x$-axis to illuminate nc-Si QDs formed inside a-Si ellipses during chemical vapour deposition,[38] and NIR TAP of the nc-Si QDs was collected from the bottom. Nanofabrication started with e-beam lithography, then using the e-beam resist of HSQ as an etching mask and the inductively coupled plasma reactive ion etching (ICP-RIE) to form nanopatterns of the a-Si metasurface. Radio frequency (RF) sputtering was used to deposit a 130 nm-thick layer of $Sb_2S_3$ on the a-Si metasurface as the tunable material, and atomic layer deposition used to deposit 10 nm-thick $Al_2O_3$ to encapsulate the device. More details on the nanofabrication process are provided in the Methods Section and Figure S2 in Supporting Information (SI).

Using finite difference time domain (FDTD) simulations, we first optimized several geometric parameters of the hybrid metasurfaces, such as the periods along $x$ and $y$ directions ($P_x$, $P_y$ respectively) and the long radius of the ellipse ($L$), to improve the interference between resonant optical modes excited within the ellipse to form dual BIC resonances. Then, we kept those parameters constant but varied the gap ($g_y$) along the *y-axis* and the tilt angle ($\theta$) of the ellipses to control the light-matter interaction inside hybrid metasurfaces. Details on the FDTD optical simulations can be found in the Methods Section and Section 1 of SI.

Symmetry breaking is induced by oppositely tilting the two elliptical pillars in a unit cell by an angle $\theta$ with respect to the *y-axis* as shown in Fig. 1b. Such symmetry breaking enables a high *Q*-factor quasi-BIC resonance in the lattice structure.[39-41] Without tilting, i.e., ($\theta$ = 0), the transmission of the hybrid metasurface at the normal incidence is high without any presence of resonances. That results from the decoupling of the true BIC resonances from the radiation continuum, i.e., without any leaky radiation modes. After titling both pillars with $\theta$ = 5°, two quasi-BIC resonant modes appear in the radiation continuum. According to the origin of materials, we named these two BICs $BIC_{a-Sb_2S_3}$ and $BIC_{a-Si}$, respectively. It can be observed that $BIC_{a-Sb_2S_3}$ was formed inside the top $Sb_2S_3$ ellipse at a wavelength of 729 nm with a *Q*-factor of 227 while $BIC_{a-Si}$ formed inside the bottom a-Si ellipse at a wavelength of 787 nm with a *Q*-factor of 373.

$Sb_2S_3$ has a huge refractive index contrast ($\Delta n$=1.2) between its amorphous and crystalline phases.[25] Simulations show that upon changing the phase of $Sb_2S_3$ from the amorphous state to crystalline in the hybrid metasurface, due to an increase in the refractive index, the wavelengths of $BIC_{a-Sb_2S_3}$ and $BIC_{a-Si}$ redshift by 45 nm and 59 nm, respectively, while the corresponding *Q*-factors are 284 and 385 respectively, as shown in Fig. 1c.

The measured optical transmission of the hybrid metasurface is shown in Fig. 1d. At the amorphous state of $Sb_2S_3$, we observed dual BIC resonances as predicted by simulations in Fig. 1c. The measured BIC resonances $BIC_{a-Sb_2S_3}$ and $BIC_{a-Si}$ have $Q$-factors of 168 and 225 at the wavelengths of 771 nm and 787 nm, respectively. The spectral positions of the measured dual BIC resonances were slightly shifted compared with the simulation results, as the simulation did not include the nonuniformity in the sputtered $Sb_2S_3$ film on the a-Si metasurface sidewalls due to the incline angle of the sputtering target with respect to the metasurface during sputtering.[42] After thermally tuning the phase of $Sb_2S_3$ from the amorphous state to crystalline, we measured the optical transmission again. We observed a redshift 46 nm and 60 nm in the dual BIC resonances of $BIC_{c-Sb_2S_3}$ and $BIC_{a-Si}$, respectively, which are consistent with the simulation results. The measured $Q$-factors of the dual BIC resonances $BIC_{c-Sb_2S_3}$ and $BIC_{a-Si}$ are 90 and 74 respectively. The degradation in the measured $Q$-factors compared with the simulations was due to the scattering losses at the rough side walls and interfaces between stacked materials.[43] The inset of Fig. 1d shows the scanning electron microscope (SEM) image of a single meta-atom of the dual-BIC hybrid metasurfaces.

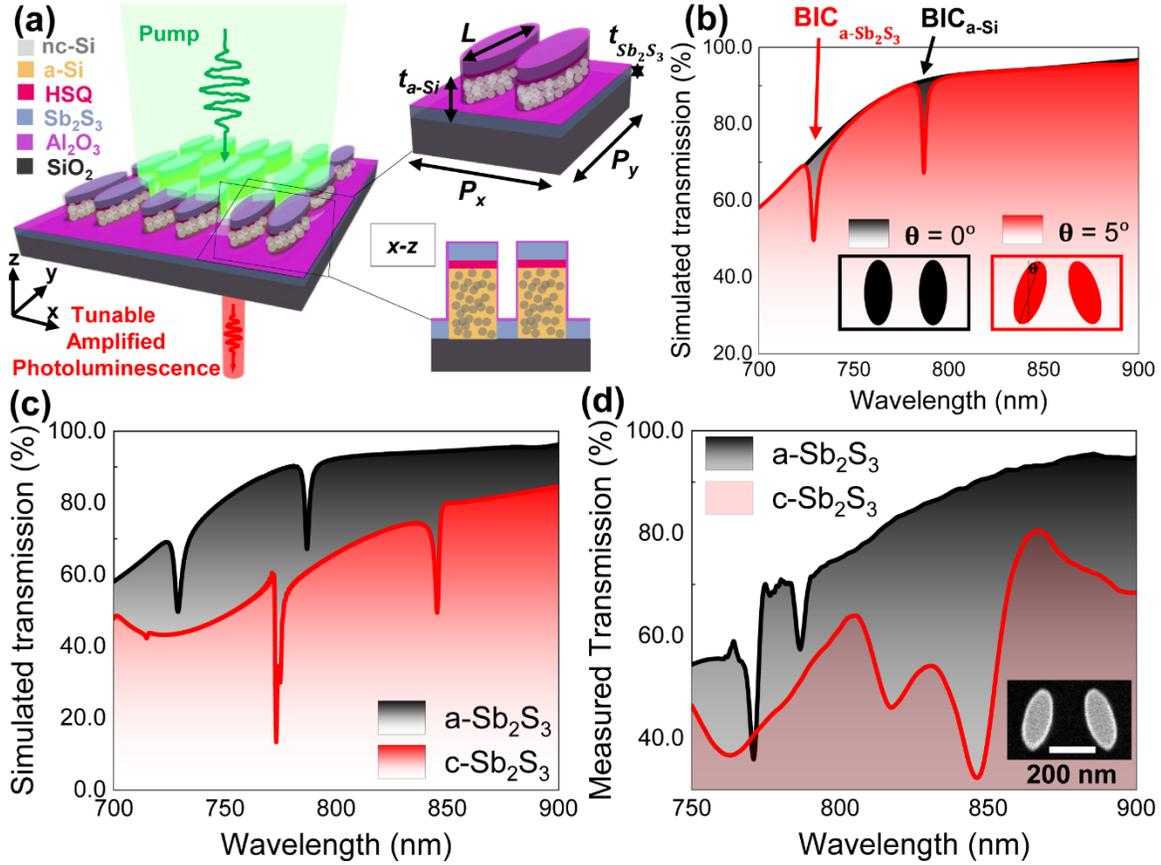

**Figure 1.** The concept of tunable amplified photoluminescence via dual BIC resonances in a hybrid metasurface in the NIR regime. (a) Schematic of the hybrid metasurface, consisting of a-Si and $Sb_2S_3$ on quartz substrate, showing the green pump laser of wavelength 532 nm and the transmitted tunable amplified photoluminescence from nc-Si QDs. (Inset) Figure shows the geometric parameters of the hybrid unit cell. Constant parameters are x-direction period ($P_x$) = 405 nm, y-direction ($P_y$) = 234 nm, thicknesses of a-Si ($t_{a-Si}$), $Sb_2S_3$ ($t_{Sb_2S_3}$), and $Al_2O_3$ are 400 nm, 130 nm, and 10 nm respectively, while the quartz substrate thickness is 500 um. The variable parameters are the long radius of ellipse (L) and vertical gap ($g_y$) and equal $g_y = L - P_y$, and the tilt angle ($\theta$) of elliptical pillars with respect to the vertical y-axis. (b) Simulated transmission without symmetry breaking ($\theta = 0°$) and with symmetry breaking ($\theta = 5°$). (c) Simulated transmission at the amorphous and crystalline states of $Sb_2S_3$ (a-$Sb_2S_3$ and c-$Sb_2S_3$ respectively) with symmetry breaking ($\theta = 5°$). (d) Measured transmission at a-$Sb_2S_3$ and c-$Sb_2S_3$ states with symmetry breaking ($\theta = 5°$). (Inset) SEM image of the fabricated hybrid metasurface with a scale bar of 200 nm.

Multipolar decomposition (MPD) calculations were carried out to investigate the interference between different optical resonant modes inside the hybrid metasurface at the BIC wavelengths as shown in Figure 2. At the amorphous state of $Sb_2S_3$, the MPD calculations are plotted in Figure 2a, indicating that the dual BICs ($BIC_{a-Sb_2S_3}$ and $BIC_{a-Si}$) arise from the coupling between the out-of-plane magnetic dipole (*MD*) mode and in-plane electric quadrupole (*EQ*) mode at both BICs wavelengths. More details about the MPD simulations are provided in the Methods Section. The electric field intensity profiles were simulated to study the field confinement inside both materials (a-Si and $Sb_2S_3$) at the dual BIC wavelengths. Figure 2b shows the field intensity distribution along the horizontal (*x-y*) plane across the middle height of the $Sb_2S_3$ and a-Si pillars respectively. In addition, the field distribution along the vertical (*x-z*) plane across the hybrid pillars. At the $BIC_{a-Sb_2S_3}$ wavelength, the electric field intensity is strongly confined inside the top $Sb_2S_3$ ellipse mainly via the *EQ* optical mode, and the bottom a-Si ellipse show *ED* and *EQ* field confinement modes. At the $BIC_{a-Si}$ wavelength, the field is confined more inside the bottom a-Si ellipse through the *EQ* optical mode compared with the top $Sb_2S_3$ ellipse. We used the vector field analysis to determine the polarity and direction of the fields inside the pillars.

In the case of c-$Sb_2S_3$, the MPD simulations plotted in Fig. 2c with a redshift in the dual BIC wavelengths are consistent with the transmission simulations. The optical modes of the shifted $BIC_{c-Sb_2S_3}$ and $BIC_{a-Si}$ are not changed compared with the amorphous state, showing the possibility of having a continuously tunable dual BIC resonances across a broad wavelength range in the NIR.[44] We plot the electric field intensity distribution inside a-Si and $Sb_2S_3$ at the shifted dual BIC wavelengths in Figure 2d. At the $BIC_{c-Sb_2S_3}$ wavelength, the intensity of the *EQ* mode inside both materials are largely confined near ellipse edges compared to the case of the amorphous state with an opposite polarity of the electric field. This is owing to the large refractive index contrast between c-$Sb_2S_3$ and a-$Sb_2S_3$, and the significantly decreased optical losses of a-Si above the wavelength of 800 nm. Similarly, at the $BIC_{a-Si}$ wavelength, the field confinement density of *EQ* mode is enhanced, leading to the improved *Q*-factor of BIC resonance. The field confinement in the *x-z* plane is strongly localized inside top c-Sb₂S₃ ellipse at the $BIC_{c-Sb_2S_3}$ and inside a-Si ellipse at the $BIC_{a-Si}$.

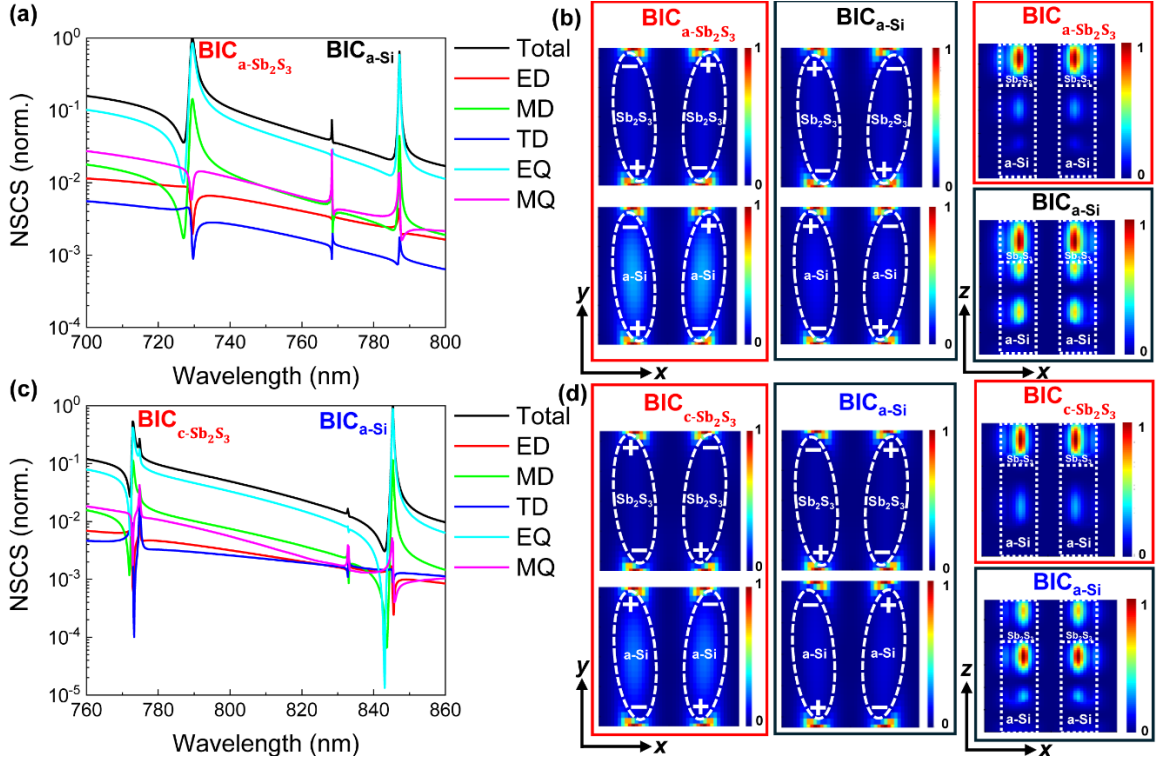

**Figure 2.** Multipolar decomposition and electric field simulations of the proposed hybrid metasurface cavity. (a) The normalized scattering cross section (NSCS) of the dual-BIC hybrid metasurfaces ($\theta$ = 5°) at the amorphous state of Sb₂S₃, showing the electric, magnetic, and toroidal modes inside the hybrid metasurface, such as the electric dipole (*ED*), electric quadrupole (*EQ*), magnetic dipole (*MD*), magnetic quadrupole (*MQ*), and toroidal dipole (*TD*). (b) The electric field intensity distribution ($|E|^2$) at a horizontal plane (*x-y*) passing through the middle height of the a-Sb₂S₃ pillars and a-Si pillars at BIC resonances formed inside a-Sb₂S₃ ($BIC_{a-Sb_2S_3}$) and inside a-Si ($BIC_{a-Si}$) respectively. Also, the electric field intensity distribution ($|E|^2$) at a vertical plane (*x-z*) through the hybrid pillars. (c) NSCS versus the wavelength of the dual-BIC hybrid metasurface ($\theta$ = 5°) at c-Sb₂S₃. (d) The electric field intensity distribution ($|E|^2$) at a horizontal plane (*x-y*) passing through the middle height of the c-Sb₂S₃ pillars and a-Si pillars at BIC resonances excited inside c-Sb₂S₃ ($BIC_{c-Sb_2S_3}$) and inside a-Si ($BIC_{a-Si}$) respectively. The electric field intensity distribution ($|E|^2$) at a vertical plane (*x-z*) through the hybrid pillars.

Besides changing the optical properties of $Sb_2S_3$ to tune the dual BIC resonances, we performed dimensional sweep of gap ($g_y$) along the *y*-direction between neighboring meta-atoms at the amorphous state of $Sb_2S_3$, illustrated in Fig. 3. The measured transmissions of different $g_y$ are plotted in Fig. 3a. Increasing $g_y$ between meta-atoms reduced the ellipse long radius and resulted in a blueshift of BIC resonance wavelength from $\lambda = 857$ nm at $g_y$ of 35 nm to $\lambda = 752$ nm at $g_y$ of 65 nm. Thus, a broadband tuning range of the BIC wavelength of up to 105 nm was achieved, which has been utilized before for spectroscopy applications using multiple arrays of BIC metasurfaces with different dimensions.[45] Corresponding SEM images of the fabricated devices are shown in Fig. 3b.

The amplified PL emissions of the fabricated patterns are characterized by a customized optical setup. The measured results of the PL enhancement are shown in Fig. 3c. The optical setup is sketched in Fig. S4 in the SI. More details about the PL measurement settings are provided in the Methods Section. We first studied the PL emission of the nc-Si QDs embedded inside an a-Si thin film prepared by chemical vapour deposition (CVD) at different deposition temperatures.[38, 46] The nc-Si QDs showed the highest PL efficiency and the largest broadband emission at a deposition temperature of 150 °C, as shown in Fig. S3 in SI. The fabricated patterns were deposited via CVD at a temperature of 150 °C. The measured PL spectrum of each fabricated pattern is plotted in Fig. 3c, and a strong PL enhancement is observed at the $BIC_{a-Si}$ wavelength which is consistent with the transmission spectrum in Fig. 3a. The normalized PL intensity shows the highest amplification of up to 15 times compared with a plain a-Si thin film using $g_y$ of 55 nm. Other dimensions give a PL enhancement factor varying from 4 to 12, as the measured *Q*-factor depends on the $g_y$ values.

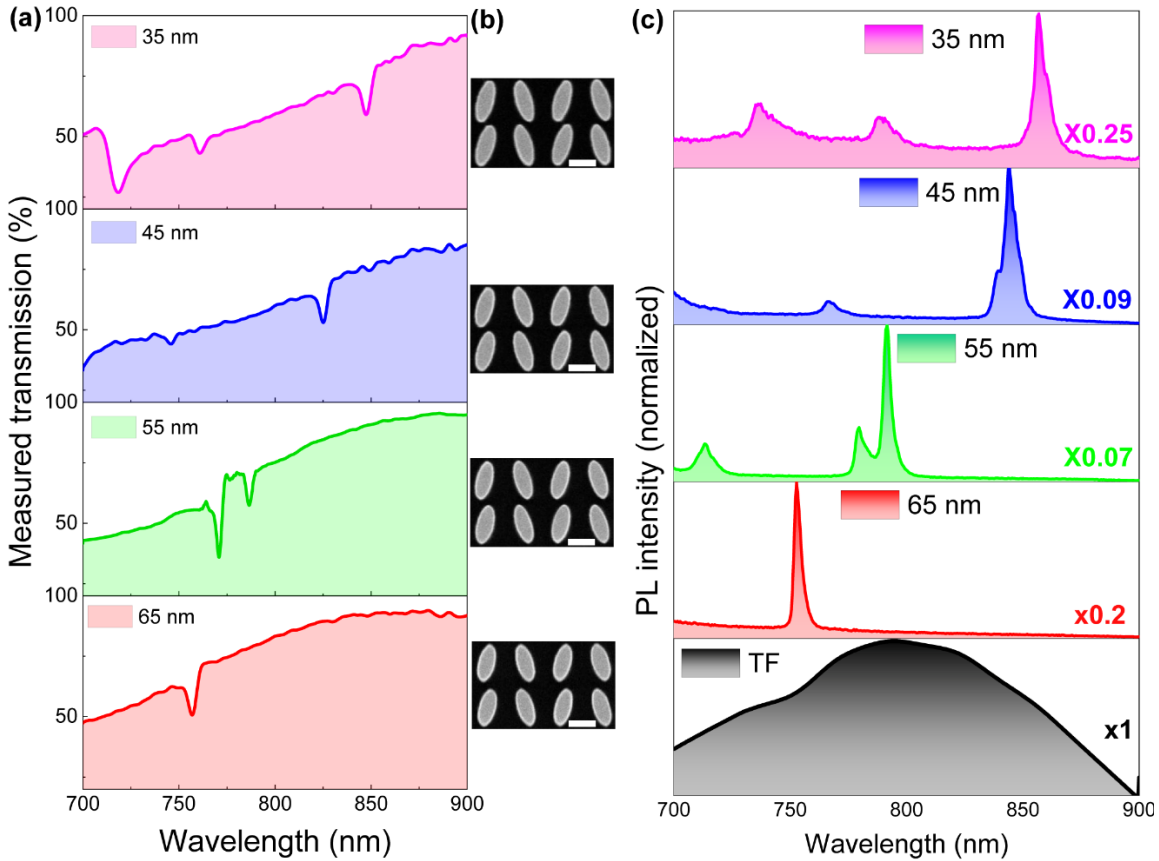

**Figure 3.** Performance of dual-BIC hybrid metasurfaces versus gap ($g_y$). (a) Measured transmission of the fabricated hybrid metasurfaces of a-$Sb_2S_3$ with different $g_y$ varying from 35 nm to 65 nm. (b) Corresponding SEM images of the fabricated hybrid metasurfaces versus different $g_y$; scale bar of 200 nm. (c) Measured normalized amplified photoluminescence (PL) spectra at different $g_y$ compared with PL of a plain a-Si thin film of a thickness of 400 nm.

TAP optical measurement for the highest $Q$-factor metasurface cavity (i.e., $g_y = 55$ nm) is plotted in Figure 4. At the amorphous state of $Sb_2S_3$, the hybrid metasurface has a peak emission at a wavelength of 791 nm with a full-width half maximum (FWHM) of 3.8 nm. After thermally changing the phase of $Sb_2S_3$, the TAP peak emission wavelength redshifts to 815 nm and the FWHM is 5.3 nm as shown in Fig. 4a. Thus, a broad tunable emission wavelength range of 24 nm is obtained through changing the optical properties of $Sb_2S_3$, which shows the feasibility of tuning and amplifying the PL after device fabrication using external stimuli rather than dimensional sweep.

In addition, we measured the TAP intensity as a function of the pump laser polarization using the same optical setup. The results are shown in Fig. 4b. The polarization of the pump laser was varied using a half-wave plate. The maximum TAP signal was obtained for the horizontal pump polarization (i.e., electric field along the *x*-direction), as the dual BICs were excited using the *x*-polarized pump laser as shown in Fig. 1b. The TAP signal vanished at the vertical pump polarization (i.e., the electric field along the *y*-direction) showing that the measured PL was enhanced mainly due to the strong light-matter interaction of BICs with nc-Si QDs inside the a-Si pillars.

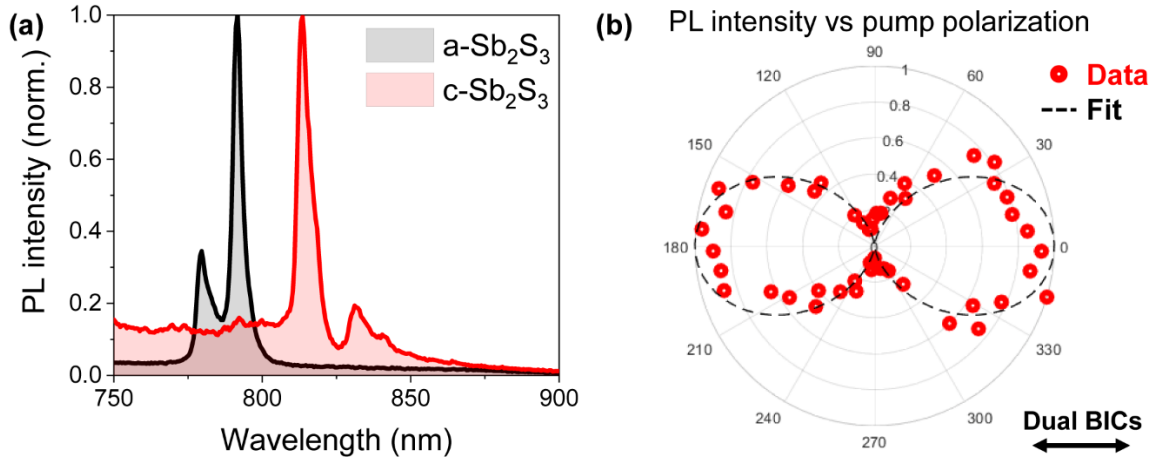

**Figure 4.** Tunable amplified photoluminescence in the proposed dual-BIC hybrid metasurfaces. (a) Normalized amplified photoluminescence (PL) in NIR regime at amorphous (a-$Sb_2S_3$) and crystalline (c-$Sb_2S_3$) states of the dual-BIC hybrid metasurfaces. (b) Polar plot of the normalized amplified photoluminescence versus the different pump polarization angle. (Inset) polarization direction of the dual BICs resonances with respect to the hybrid metasurface direction.

Finally, by engineering the phase gradient of dual-BIC hybrid metasurfaces we designed a tunable metalens that amplify PL emission through supporting dual BICs high *Q*-factor with similar values as hybrid metasurface. We focused the tunable amplified PL (FTAP) from the hybrid dual BICs using metalens as shown in Fig. 5a. We constructed the phase profile of the high *Q*-factor cavity through rotating the meta-atom from 0 to 360° around *z*-axis at the three phases of $Sb_2S_3$, i.e., amorphous (a-$Sb_2S_3$), semi-crystalline (i-$Sb_2S_3$), and crystalline (c-$Sb_2S_3$), as shown in Fig. S5 in SI. The transmission efficiency of the designed metalens is more than 65% for all $Sb_2S_3$ phases. At the amorphous state of $Sb_2S_3$, shown in Fig. 5b, we plotted the intensity at the focal plane at three wavelengths (791 nm, 803 nm, and 815 nm), which corresponded to the BIC resonances at the three phases respectively. We observed a strong FTAP signal at the wavelength of 791 nm with a diffraction-limited focused spot of FWHM down to 0.49 um. At the wavelengths of 803 nm and 815 nm, we achieved relatively weak focusing due to the focusing of nonamplified PL emission of the nc-Si QDs at these wavelengths. The right panel of Fig. 5b shows the normalized electric field distribution in the *x-y* plane at the focal plane. In the case of semi-crystalline $Sb_2S_3$ in Fig. 5b, the normalized intensity at the focal point was maximum at the wavelength of 803 nm which is the wavelength of BIC resonance of hybrid metasurface using semi-crystalline $Sb_2S_3$ phase. The relative focusing efficiency at the wavelength of 791 nm and 815 nm were below 0.6% as the PL emission was not amplified at these wavelengths in this case. The two-dimensional (2D) field profile shows strong focused light spot of FWHM equal 0.47 um at the wavelength of 803 nm. At the full crystalline phase of $Sb_2S_3$ in Fig. 5b, the maximum focusing efficiency is acquired at the wavelength of 815 nm with a focused spot of FWHM equal 0.47 um. The field profile shows a bright normalized FTAP signal at the wavelength of 815 nm and a weak amplifying focusing efficiency below 0.8% at the wavelengths of 791 nm and 803 nm, respectively.

Furthermore, it is worth noting that, we used a tunable optical stimulus to realize intermediate multi-states in $Sb_2S_3$ as shown in Fig. 5c. Intermediate states of $Sb_2S_3$ requires obtaining a $Sb_2S_3$ film with a mixture of amorphous and crystalline grains with various percentages.[47] In this study, by gradually increasing the power density of the incident laser ($\lambda$ = 532 nm, CW) up to 200 kW/cm$^2$ on the $Sb_2S_3$ film, the Raman shift of a-$Sb_2S_3$ gradually changes to several intermediate states i-$Sb_2S_3$(n) till c-$Sb_2S_3$. By exposure the $Sb_2S_3$ film with a certain laser power density, we can get arbitrary percentage of the crystalline grains in the amorphous film till getting the desired phase of $Sb_2S_3$. This could be useful for future multi-functional reconfigurable metasurfaces.

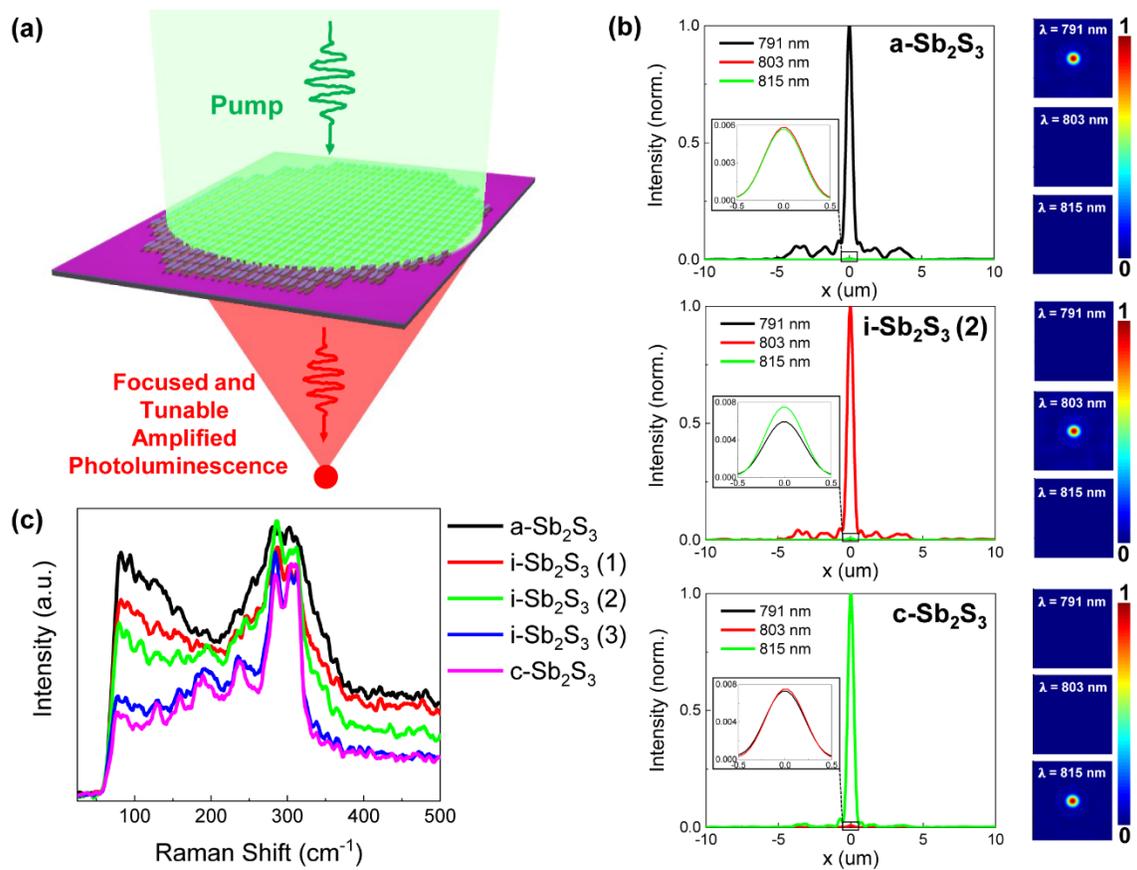

**Figure 5.** Proposed concept for focused and tunable amplified photoluminescence (FTAP) in tunable hybrid metalens supporting BIC resonances. (a) 3D schematic of the hybrid metalens with the pump laser at 532 nm wavelength and FTAP in transmission mode at NIR regime. (b) Simulated normalized intensity of FTAP at the focal plane of the hybrid metalens. Simulations were performed at the three wavelengths that correspond to the BIC resonances wavelength, i.e., 791 nm, 803 nm, and 815 nm and at three different $Sb_2S_3$ phases for amorphous (a- $Sb_2S_3$), intermediate (i-$Sb_2S_3$ (2)), and crystalline (c-$Sb_2S_3$) phases, respectively. (Inset) simulated 2D electric field intensity ($|E|^2$) plot of the *x-y* plane at the focal point for the wavelengths of 791 nm, 803 nm, and 815 nm, respectively at the corresponding $Sb_2S_3$ phase. (c) Measured Raman shift signal for $Sb_2S_3$ thin film after CW-laser annealing showing a gradual change from an amorphous phase (a-$Sb_2S_3$) to several intermediate phases (i-$Sb_2S_3$ (n)) with different crystallinity ratios to a full crystalline phase (c-$Sb_2S_3$).

**CONCLUSION**

In this work, we have demonstrated a hybrid metasurface device that enables FTAP emission by combining nc-Si QDs embedded inside a-Si metasurface and $Sb_2S_3$ as a modulator. The hybrid metasurface exploits dual BICs in the NIR to achieve significant PL amplification and dynamic tunability. By integrating high-temperature-stable nc-Si QDs, we address the challenge of reliable phase transitions in PCMs while mitigating the temperature stability issues of other semiconductor nanocrystals. The system achieves up to 15-fold PL enhancement with a tunable wavelength shift of up to 105 nm through dimensional modulation, while the phase transition of $Sb_2S_3$ enables all-optical tunable emission across a spectral range of 24 nm. Moreover, the introduction of a high-*Q*-factor metalens enables a diffraction-limited focused light source in the NIR spectrum. This hybrid metasurface-based approach paves the way for reconfigurable nanophotonic devices, offering new possibilities for highly efficient quantum light sources in next-generation photonic systems. Our work highlights the potential of hybrid metasurfaces in advancing quantum technologies, optical communications, solar energy and advanced sensing applications,[48-66] providing a scalable solution for tunable and amplified light emission in integrated photonic devices.[67]

## METHODS

**Transmission and Field Profile Simulations**

Linear optical simulations were conducted using a three-dimensional finite-difference time-domain (FDTD) simulator (Lumerical FDTD Solutions).[68] The refractive indices of the materials, shown in Figure S1 in SI, were extracted from ellipsometry measurements after material deposition. A linearly polarized (*x*-direction) plane wave was employed to excite the hybrid metasurface cavity, with periodic boundary conditions (PBC) applied along the horizontal (*x*-*y*) plane. A perfectly matched layer (PML) was used in the vertical (*z*-axis) direction to prevent spurious reflections at the simulation boundaries. The mesh size was set to 10 nm or smaller to ensure adequate resolution. A field monitor was positioned beneath the quartz substrate to measure light transmission, and two horizontal 2D field monitor was placed at the middle height of the a-Si and $Sb_2S_3$ elliptical pillars to capture the field profile at BIC resonances. An auto-shutoff criterion of $5 \times 10^{-7}$ was employed to ensure high-accuracy simulations.

**Multipolar decomposition calculations**

Multipolar decomposition calculations were conducted using the Lumerical FDTD simulator. Two 3D field monitors were positioned around the hybrid metasurface cavity. The first monitor computed the electric and magnetic field components confined within the metasurface nanostructures at each mesh point, while the second monitor determined the effective refractive index across the cavity volume. The electric and magnetic scattering cross-sections were then calculated using the following equations:

$$C_{ED} = \frac{k_0^4}{6\pi\epsilon_0^2 E_0^2} \left| p_{car} + \frac{ik_0}{c}\left(t + \frac{k_0^2}{10}\overline{R_t^2}\right)\right|^2 \quad (2),$$

$$C_{EQ} = \frac{k_0^6}{80\pi\epsilon_0^2 E_0^2} \left| \overline{\overline{Q_e}} + \frac{ik_0}{c}\overline{\overline{Q_t}}\right|^2 \quad (3),$$

$$C_{MD} = \frac{\eta_0^2 k_0^4}{6\pi E_0^2} \left| m_{car} - k_0^2 \overline{R_m^2} \right|^2 \qquad (4),$$

$$C_{MQ} = \frac{\eta_0^2 k_0^6}{80\pi E_0^2} \left| \overline{\overline{Q_m}} \right|^2 \qquad (5),$$

where $C_{ED}$, $C_{EQ}$, $C_{MD}$, and $C_{MQ}$ denote the scattering cross-section of an electric dipole, electric quadrupole, magnetic dipole, and magnetic quadrupole, respectively. $p_{car}$, $t$, $m_{car}$, $\overline{\overline{Q_e}}$, and $\overline{\overline{Q_m}}$ are scattering power components and refers to the electric, toroidal, and magnetic dipoles, electric and magnetic quadrupole, respectively. The detailed equations for these components are in the literature.[69]

**Fabrication of hybrid BICs metasurface**

Figure S2 shows the fabrication flow of the a-Si/Sb$_2$S$_3$ hybrid metasurface. Transparent deep ultraviolet quartz substrate (Photonik Singapore) was cleaned using an ultrasonic bath at room temperature using acetone and IPA. Next, an a-Si thin film of thickness 400 nm was deposited using inductively coupled plasma chemical vapour deposition (ICP-CVD, Oxford Plasmalab System 380) at the substrate temperature varying from 50 °C to 250 °C, using SiH$_4$ and Ar gases at a flow rate of 45 sccm and 30 sccm respectively. ICP-CVD chamber pressure was 8 mTorr, and the plasma used had RF power of 50 W and ICP power of 3000 W.

The third fabrication step was electron beam lithography (EBL, Elionix ELS-7000). The e-beam resist, hydrogen silsesquioxane (HSQ), was spin coated at 2000 rpm for 60 sec, followed by an anti-charging resist (EZspacer), spin-coated at 1500 rpm for 30 sec to reduce EBL exposure error caused by charging effect. Exposure was done using an e-beam current of 500 pA and accelerated with a voltage bias of 100 kV. The real metasurface area size was 100 x 100 μm$^2$ in a field pattern of 300 x 300 μm$^2$ to increase the metasurface array effect. The base dose charge density was 9600 μc/cm$^2$, and the field area was digitized into 60,000 dots. Development of HSQ after EBL exposure started with removing EZspacer by rinsing the sample in deionized (DI) water for a few seconds. Then, the sample was soaked in a salty developer (1% NaOH, 4% NaCl in DI water) for 60 sec, then soaked in fresh DI water for another 60 sec to stop the development of HSQ. Sample was cleaned after development by rinsing it with DI water and IPA for a few seconds and then dried using N$_2$ gas flow.

Transferring HSQ nanopatterns to a-Si was done using inductively coupled plasma reactive ion etching (ICP-RIE, Oxford OIPT Plasmalab system). Using pure Cl$_2$ gas at a flow rate of 22 sccm and chamber pressure of 5 mTorr at room temperature, Cl$_2$ plasma was ignited using RF power 200 W and ICP power 400 W. Without removing the remaining HSQ mask of thickness ~50 nm, we deposited Sb$_2$S$_3$ of a thickness of 130 nm using RF sputtering (UBM) at room temperature to avoid phase change for Sb$_2$S$_3$. Ar plasma used had RF power 20 W and gas flow 21.6 sccm at chamber pressure 10 mTorr.

The last fabrication step was adding a protection layer Al$_2$O$_3$ of thickness 10 nm using atomic layer deposition (ALD, Beneq TFS 200). Using trimethylaluminium (TMA) and water (H$_2$O) precursors, reactions were run for 100 cycles at a chamber pressure of 4 mTorr and a substrate temperature of 80 °C. N$_2$ was used to purge remaining gas precursors in between cycles. ALD deposition done at substrate temperature of 80 °C.

**Transmission measurement**

Transmission measurements were carried out using a supercontinuum nanosecond laser (Opera, Leukos Laser Inc.) operating at a repetition rate of 30 kHz. To generate x-polarized light for exciting the dual BIC metasurface, a quarter-wave plate (QWP), a half-wave plate (HWP), and a linear polarizer (LP) were employed in the setup. The transmitted light was collected and coupled into an optical fiber, which was then directed to an Ocean Optics USB4000 spectrometer for analysis. The spectrometer was connected to a computer for data acquisition.

**PL and polarization measurement**

The optical setup used in all PL measurements of the resonant metasurface cavities is shown in Figure S4. We used a fiber-coupled continuous-wave (CW) laser at a wavelength of 532 nm (WiTEC Alpha300), with an average power up to 50 mW. A rotational LP was used to measure the PL emission at different pump polarization angles. A dichoric mirror was used to reflect the 532 nm laser to a 5x objective with a numerical aperture of 0.15 and a diffraction-limited spot diameter of ~ 2.2 μm. Sample was mounted on a motorized XYZ-stage to allow 2D PL scan and $z$-focusing. The PL was collected using the same 5x objective. A notch filter of wavelength 532 nm was used to block the pump signal from the PL and protect the PL detector. White light was used to align the CW laser on the metasurface pattern using a beam splitter (BS). A flipping mirror (M2) was used to guide the reflected white light from the sample to the focusing lens (L1) and CCD camera for imaging purposes. The WiTEC PL detector was used to measure the PL spectrum, with a grating of 150 or 600 grooves/mm. A linear polarizer was placed before the PL detector to resolve the PL enhancement versus polarization angle.

**ASSOCIATED CONTENT**

**Supporting Information**

Measured refractive indices of a-Si, $Sb_2S_3$, and $Al_2O_3$; nanofabrication flow for the hybrid metasurface cavity; PL of nc-Si QDs at different CVD deposition temperatures; PL measurement setup; Phase vs rotation for a-$Sb_2S_3$, i-$Sb_2S_3$, and c-$Sb_2S_3$ metasurfaces and transmission; comparison table with literature works for PL amplification, tunable PL, focusing of PL, and Q-factor.

## AUTHOR INFORMATION

### Corresponding Authors

**Omar A. M. Abdelraouf** – Institute of Materials Research and Engineering, Agency for Science, Technology, and Research (A*STAR), 2 Fusionopolis Way, #08-03, Innovis, Singapore 138634, Singapore; orcid.org/0000-0002-9065-7414; Email: omar_abdelrahman@imre.a-star.edu.sg

**Hong Liu** – Institute of Materials Research and Engineering, Agency for Science, Technology, and Research (A*STAR), 2 Fusionopolis Way, #08-03, Innovis, Singapore 138634, Singapore; orcid.org/0000-0002-3560-9401; Email: h-liu@imre.a-star.edu.sg

### Authors

**Mengfei Wu** – Institute of Materials Research and Engineering, Agency for Science, Technology, and Research (A*STAR), 2 Fusionopolis Way, #08-03, Innovis, Singapore 138634, Singapore; Department of Materials Science and Engineering, National University of Singapore, 9 Engineering Drive 1, #03-09 EA, Singapore 117575, Singapore; orcid.org/0000- 0002-4728-3234

## AUTHOR CONTRIBUTIONS

O.A.M.A. conceived the idea, designed the multi-BIC cavity and tunable metalens, carried out FDTD simulations and multipolar decomposition, performed materials deposition and characterization, nanofabrication of the dual BIC cavity, linear optical measurements, photoluminescence measurements, optical tune phase change materials, and wrote the original draft of the manuscript. M.W. assisted with optical measurements and manuscript revision. L.H. acquired funding and supervised the project.

**Competing Financial Interests**

The authors declare no competing financial interest.

**ACKNOWLEDGMENT**

L.H. acknowledge the support from RIE2025 Manufacturing, Trade and Connectivity (MTC) Programmatic Fund (M23M2b0056). M.W. acknowledge the Manufacturing, Trade and Connectivity (MTC) Programmatic Fund (M21J9b0085). O.A.M.A. acknowledge the hands-on support of Aravind P Anthur during building the optical setup and the scientific discussion with Prof. Wang Qijie and Prof. Wang Xiao Renshaw.

**REFERENCES**

(1) Pavesi, L.; Dal Negro, L.; Mazzoleni, C.; Franzo, G.; Priolo, d. F. Optical gain in silicon nanocrystals. *Nature* **2000**, *408* (6811), 440-444.

(2) Sain, B.; Das, D. Tunable photoluminescence from nc-Si/a-SiN x: H quantum dot thin films prepared by ICP-CVD. *Physical Chemistry Chemical Physics* **2013**, *15* (11), 3881-3888.

(3) Sangghaleh, F.; Sychugov, I.; Yang, Z.; Veinot, J. G.; Linnros, J. Near-unity internal quantum efficiency of luminescent silicon nanocrystals with ligand passivation. *ACS nano* **2015**, *9* (7), 7097-7104.

(4) Dohnalová, K.; Poddubny, A. N.; Prokofiev, A. A.; De Boer, W. D.; Umesh, C. P.; Paulusse, J. M.; Zuilhof, H.; Gregorkiewicz, T. Surface brightens up Si quantum dots: direct bandgap-like size-tunable emission. *Light: science & applications* **2013**, *2* (1), e47-e47.

(5) Sychugov, I.; Valenta, J.; Linnros, J. Probing silicon quantum dots by single-dot techniques. *Nanotechnology* **2017**, *28* (7), 072002.

(6) Abdul-Ameer, N. M.; Abdulrida, M. C. Direct optical energy gap in amorphous silicon quantum dots. *Journal of Modern Physics* **2011**, *2011*.

(7) Dal Negro, L.; Cazzanelli, M.; Pavesi, L.; Ossicini, S.; Pacifici, D.; Franzò, G.; Priolo, F.; Iacona, F. Dynamics of stimulated emission in silicon nanocrystals. *Applied Physics Letters* **2003**, *82* (26), 4636-4638.

(8) Dal Negro, L.; Cazzanelli, M.; Daldosso, N.; Gaburro, Z.; Pavesi, L.; Priolo, F.; Pacifici, D.; Franzo, G.; Iacona, F. Stimulated emission in plasma-enhanced chemical vapour deposited silicon nanocrystals. *Physica E: Low-dimensional Systems and Nanostructures* **2003**, *16* (3-4), 297-308.

(9) Belyakov, V.; Burdov, V.; Lockwood, R.; Meldrum, A. Silicon nanocrystals: fundamental theory and implications for stimulated emission. *Advances in Optical Technologies* **2008**, *2008* (1), 279502.

(10) Wen, X.; Zhang, P.; Smith, T. A.; Anthony, R. J.; Kortshagen, U. R.; Yu, P.; Feng, Y.; Shrestha, S.; Coniber, G.; Huang, S. Tunability limit of photoluminescence in colloidal silicon nanocrystals. *Scientific reports* **2015**, *5* (1), 12469.

(11) Yu, Y.; Fan, G.; Fermi, A.; Mazzaro, R.; Morandi, V.; Ceroni, P.; Smilgies, D.-M.; Korgel, B. A. Size-dependent photoluminescence efficiency of silicon nanocrystal quantum dots. *The Journal of Physical Chemistry C* **2017**, *121* (41), 23240-23248.

(12) English, D. S.; Pell, L. E.; Yu, Z.; Barbara, P. F.; Korgel, B. A. Size tunable visible luminescence from individual organic monolayer stabilized silicon nanocrystal quantum dots. *Nano Letters* **2002**, *2* (7), 681-685.

(13) Paniagua-Dominguez, R.; Ha, S. T.; Kuznetsov, A. I. Active and tunable nanophotonics with dielectric nanoantennas. *Proceedings of the IEEE* **2019**, *108* (5), 749-771.

(14) Aftenieva, O.; Brunner, J.; Adnan, M.; Sarkar, S.; Fery, A.; Vaynzof, Y.; König, T. A. Directional amplified photoluminescence through large-area perovskite-based metasurfaces. *ACS nano* **2023**, *17* (3), 2399-2410.

(15) Tripathi, A.; Kruk, S.; Shang, Y.; Zhou, J.; Kravchenko, I.; Jin, D.; Kivshar, Y. Topological nanophotonics for photoluminescence control. *Nanophotonics* **2020**, *10* (1), 435-441.
(16) Vaskin, A.; Kolkowski, R.; Koenderink, A. F.; Staude, I. Light-emitting metasurfaces. *Nanophotonics* **2019**, *8* (7), 1151-1198.
(17) Kogos, L. C.; Paiella, R. Light emission near a gradient metasurface. *ACS Photonics* **2016**, *3* (2), 243-248.
(18) Liu, S.; Vaskin, A.; Addamane, S.; Leung, B.; Tsai, M.-C.; Yang, Y.; Vabishchevich, P. P.; Keeler, G. A.; Wang, G.; He, X. Light-emitting metasurfaces: simultaneous control of spontaneous emission and far-field radiation. *Nano letters* **2018**, *18* (11), 6906-6914.
(19) Muhammad, N.; Chen, Y.; Qiu, C.-W.; Wang, G. P. Optical bound states in continuum in MoS2-based metasurface for directional light emission. *Nano Letters* **2021**, *21* (2), 967-972.
(20) Zhang, Z.; Xu, C.; Liu, C.; Lang, M.; Zhang, Y.; Li, M.; Lu, W.; Chen, Z.; Wang, C.; Wang, S. Dual control of enhanced quasi-bound states in the continuum emission from resonant c-si metasurfaces. *Nano Letters* **2023**, *23* (16), 7584-7592.
(21) Li, Z.; Panmai, M.; Zhou, L.; Li, S.; Liu, S.; Zeng, J.; Lan, S. Optical sensing and switching in the visible light spectrum based on the bound states in the continuum formed in GaP metasurfaces. *Applied Surface Science* **2023**, *620*, 156779.
(22) Abdollahramezani, S.; Hemmatyar, O.; Taghinejad, H.; Krasnok, A.; Kiarashinejad, Y.; Zandehshahvar, M.; Alù, A.; Adibi, A. Tunable nanophotonics enabled by chalcogenide phase-change materials. *Nanophotonics* **2020**, *9* (5), 1189-1241.
(23) Chaudhary, K.; Tamagnone, M.; Yin, X.; Spägele, C. M.; Oscurato, S. L.; Li, J.; Persch, C.; Li, R.; Rubin, N. A.; Jauregui, L. A. Polariton nanophotonics using phase-change materials. *Nature communications* **2019**, *10* (1), 4487.
(24) Tripathi, D.; Vyas, H. S.; Kumar, S.; Panda, S. S.; Hegde, R. Recent developments in Chalcogenide phase change material-based nanophotonics. *Nanotechnology* **2023**, *34* (50), 502001.
(25) Delaney, M.; Zeimpekis, I.; Lawson, D.; Hewak, D. W.; Muskens, O. L. A new family of ultralow loss reversible phase‐change materials for photonic integrated circuits: Sb2S3 and Sb2Se3. *Advanced functional materials* **2020**, *30* (36), 2002447.
(26) Jana, S.; Sreekanth, K. V.; Abdelraouf, O. A.; Lin, R.; Liu, H.; Teng, J.; Singh, R. Aperiodic Bragg reflectors for tunable high-purity structural color based on phase change material. *Nano Letters* **2024**, *24* (13), 3922-3929.
(27) Abdelraouf, O. A.; Wang, X. C.; Goh Ken, C. H.; Lim Nelson, C. B.; Ng, S. K.; Wang, W. D.; Renshaw Wang, X.; Wang, Q. J.; Liu, H. All‐Optical Switching of Structural Color with a Fabry–Pérot Cavity. *Advanced Photonics Research* **2023**, *4* (11), 2300209.
(28) Jeon, K. A.; Kim, J. H.; Choi, J. B.; Han, K. B.; Lee, S. Y. Annealing effect on the photoluminescence of Si nanocrystallites thin films. *Materials Science and Engineering: C* **2003**, *23* (6-8), 1017-1019.
(29) Guan, J.; Park, J.-E.; Deng, S.; Tan, M. J.; Hu, J.; Odom, T. W. Light–matter interactions in hybrid material metasurfaces. *Chemical Reviews* **2022**, *122* (19), 15177-15203.

(30) Lepeshov, S. I.; Krasnok, A. E.; Belov, P. A.; Miroshnichenko, A. E. Hybrid nanophotonics. *Physics-Uspekhi* **2019**, *61* (11), 1035.
(31) Makarov, S. V.; Milichko, V.; Ushakova, E. V.; Omelyanovich, M.; Cerdan Pasaran, A.; Haroldson, R.; Balachandran, B.; Wang, H.; Hu, W.; Kivshar, Y. S. Multifold emission enhancement in nanoimprinted hybrid perovskite metasurfaces. *Acs Photonics* **2017**, *4* (4), 728-735.
(32) Choi, C.; Mun, S. E.; Sung, J.; Choi, K.; Lee, S. Y.; Lee, B. Hybrid state engineering of phase‐change metasurface for all‐optical cryptography. *Advanced Functional Materials* **2021**, *31* (4), 2007210.
(33) Cheng, Y.; Chen, F.; Luo, H. Triple-band perfect light absorber based on hybrid metasurface for sensing application. *Nanoscale research letters* **2020**, *15*, 1-10.
(34) Wang, Z.; Dong, Y.; Peng, Z.; Hong, W. Hybrid metasurface, dielectric resonator, low-cost, wide-angle beam-scanning antenna for 5G base station application. *IEEE Transactions on Antennas and Propagation* **2022**, *70* (9), 7646-7658.
(35) Solntsev, A. S.; Agarwal, G. S.; Kivshar, Y. S. Metasurfaces for quantum photonics. *Nature Photonics* **2021**, *15* (5), 327-336.
(36) Guo, M.; Huang, L.; Liu, W.; Ding, J. Hybrid metasurface comprising epsilon-near-zero material for double transparent windows in optical communication band. *Optical Materials* **2021**, *112*, 110802.
(37) Qin, J.; Jiang, S.; Wang, Z.; Cheng, X.; Li, B.; Shi, Y.; Tsai, D. P.; Liu, A. Q.; Huang, W.; Zhu, W. Metasurface micro/nano-optical sensors: principles and applications. *ACS nano* **2022**, *16* (8), 11598-11618.
(38) Hernandez, A. V.; Torchynska, T.; Vazquez, A. Q.; Matsumoto, Y.; Khomenkova, L.; Shcherbina, L. Emission and structure investigations of Si nano-crystals embedded in amorphous silicon. In *Journal of Physics: Conference Series*, 2007; IOP Publishing: Vol. 61, p 1231.
(39) Abdelraouf, O. A.; Anthur, A. P.; Wang, X. R.; Wang, Q. J.; Liu, H. Modal phase-matched bound states in the continuum for enhancing third harmonic generation of deep ultraviolet emission. *ACS nano* **2024**, *18* (5), 4388-4397.
(40) Azzam, S. I.; Kildishev, A. V. Photonic bound states in the continuum: from basics to applications. *Advanced Optical Materials* **2021**, *9* (1), 2001469.
(41) Moretti, G. Q.; Tittl, A.; Cortés, E.; Maier, S. A.; Bragas, A. V.; Grinblat, G. Introducing a Symmetry‐Breaking Coupler into a Dielectric Metasurface Enables Robust High‐Q Quasi‐BICs. *Advanced Photonics Research* **2022**, *3* (12), 2200111.
(42) Gago, R.; Vázquez, L.; Cuerno, R.; Varela, M.; Ballesteros, C.; Albella, J. M. Nanopatterning of silicon surfaces by low-energy ion-beam sputtering: dependence on the angle of ion incidence. *Nanotechnology* **2002**, *13* (3), 304.
(43) Bennett, J. Optical scattering and absorption losses at interfaces and in thin films. *Thin Solid Films* **1985**, *123* (1), 27-44.
(44) Abdelraouf, O. A.; Anthur, A. P.; Dong, Z.; Liu, H.; Wang, Q.; Krivitsky, L.; Renshaw Wang, X.; Wang, Q. J.; Liu, H. Multistate tuning of third harmonic generation in fano‐resonant hybrid dielectric metasurfaces. *Advanced Functional Materials* **2021**, *31* (48), 2104627.

(45) Yesilkoy, F.; Arvelo, E. R.; Jahani, Y.; Liu, M.; Tittl, A.; Cevher, V.; Kivshar, Y.; Altug, H. Ultrasensitive hyperspectral imaging and biodetection enabled by dielectric metasurfaces. *Nature Photonics* **2019**, *13* (6), 390-396.

(46) Hsieh, Y.-L.; Kau, L.-H.; Huang, H.-J.; Lee, C.-C.; Fuh, Y.-K.; Li, T. T. In situ plasma monitoring of PECVD nc-Si: H films and the influence of dilution ratio on structural evolution. *Coatings* **2018**, *8* (7), 238.

(47) Abdelraouf, O. A.; Wang, Z.; Liu, H.; Dong, Z.; Wang, Q.; Ye, M.; Wang, X. R.; Wang, Q. J.; Liu, H. Recent advances in tunable metasurfaces: materials, design, and applications. *ACS nano* **2022**, *16* (9), 13339-13369.

(48) Wang, K.; Chekhova, M.; Kivshar, Y. Metasurfaces for quantum technologies. *Physics Today* **2022**, *75* (8), 38-44.

(49) Abdelraouf, O. A.; Shaker, A.; Allam, N. K. Novel design of plasmonic and dielectric antireflection coatings to enhance the efficiency of perovskite solar cells. *Solar Energy* **2018**, *174*, 803-814.

(50) Abdelraouf, O. A.; Shaker, A.; Allam, N. K. Plasmonic nanoscatter antireflective coating for efficient CZTS solar cells. In *Photonics for Solar Energy Systems VII*, 2018; SPIE: Vol. 10688, pp 15-23.

(51) Abdelraouf, O. A.; Shaker, A.; Allam, N. K. Design of optimum back contact plasmonic nanostructures for enhancing light coupling in CZTS solar cells. In *Photonics for Solar Energy Systems VII*, 2018; SPIE: Vol. 10688, pp 33-41.

(52) Abdelraouf, O. A.; Shaker, A.; Allam, N. K. Design methodology for selecting optimum plasmonic scattering nanostructures inside CZTS solar cells. In *Photonics for Solar Energy Systems VII*, 2018; SPIE: Vol. 10688, pp 24-32.

(53) Abdelraouf, O. A.; Shaker, A.; Allam, N. K. Enhancing light absorption inside CZTS solar cells using plasmonic and dielectric wire grating metasurface. In *Metamaterials XI*, 2018; SPIE: Vol. 10671, pp 165-174.

(54) Abdelraouf, O. A.; Shaker, A.; Allam, N. K. All dielectric and plasmonic cross-grating metasurface for efficient perovskite solar cells. In *Metamaterials Xi*, 2018; SPIE: Vol. 10671, pp 104-112.

(55) Abdelraouf, O. A.; Shaker, A.; Allam, N. K. Using all dielectric and plasmonic cross grating metasurface for enhancing efficiency of CZTS solar cells. In *Nanophotonics VII*, 2018; SPIE: Vol. 10672, pp 246-255.

(56) Atef, N.; Emara, S. S.; Eissa, D. S.; El‐Sayed, A.; Abdelraouf, O. A.; Allam, N. K. Well‐dispersed Au nanoparticles prepared via magnetron sputtering on TiO2 nanotubes with unprecedentedly high activity for water splitting. *Electrochemical Science Advances* **2021**, *1* (1), e2000004.

(57) Shlezinger, N.; Alexandropoulos, G. C.; Imani, M. F.; Eldar, Y. C.; Smith, D. R. Dynamic metasurface antennas for 6G extreme massive MIMO communications. *IEEE Wireless Communications* **2021**, *28* (2), 106-113.

(58) Abdelraouf, O. A.; Shaker, A.; Allam, N. K. Front dielectric and back plasmonic wire grating for efficient light trapping in perovskite solar cells. *Optical materials* **2018**, *86*, 311-317.

(59) Abdelraouf, O. A.; Allam, N. K. Towards nanostructured perovskite solar cells with enhanced efficiency: Coupled optical and electrical modeling. *Solar Energy* **2016**, *137*, 364-370.

(60) Abdelraouf, O. A.; Allam, N. K. Nanostructuring for enhanced absorption and carrier collection in CZTS-based solar cells: coupled optical and electrical modeling. *Optical Materials* **2016**, *54*, 84-88.

(61) Abdelraouf, O. A.; Abdelrahaman, M. I.; Allam, N. K. Plasmonic scattering nanostructures for efficient light trapping in flat czts solar cells. In *Metamaterials XI*, 2017; SPIE: Vol. 10227, pp 90-98.

(62) Abdelraouf, O. A.; Ali, H. A.; Allam, N. K. Optimizing absorption and scattering cross section of metal nanostructures for enhancing light coupling inside perovskite solar cells. In *2017 Conference on Lasers and Electro-Optics Europe & European Quantum Electronics Conference (CLEO/Europe-EQEC)*, 2017; IEEE: pp 1-1.

(63) Chen, X.; Fan, W. Toroidal metasurfaces integrated with microfluidic for terahertz refractive index sensing. *Journal of Physics D: Applied Physics* **2019**, *52* (48), 485104.

(64) Liu, H.; Wang, H.; Wang, H.; Deng, J.; Ruan, Q.; Zhang, W.; Abdelraouf, O. A.; Ang, N. S. S.; Dong, Z.; Yang, J. K. High-order photonic cavity modes enabled 3D structural colors. *ACS nano* **2022**, *16* (5), 8244-8252.

(65) Abdelraouf, O. A.; Mousa, A.; Ragab, M. Next-Generation Multi-layer Metasurface Design: Hybrid Deep Learning Models for Beyond-RGB Reconfigurable Structural Colors. *arXiv preprint arXiv:2409.07121* **2024**.

(66) Abdelraouf, O. A.; Anthur, A. P.; Liu, H.; Dong, Z.; Wang, Q.; Krivitsky, L.; Wang, X. R.; Wang, Q. J.; Liu, H. Tunable transmissive THG in silicon metasurface enabled by phase change material. In *CLEO: QELS_Fundamental Science*, 2021; Optica Publishing Group: p FTh4K. 3.

(67) Wang, Q. H.; Ni, P. N.; Xie, Y. Y.; Kan, Q.; Chen, P. P.; Fu, P.; Deng, J.; Jin, T. L.; Chen, H. D.; Lee, H. W. H. On‐chip generation of structured light based on metasurface optoelectronic integration. *Laser & Photonics Reviews* **2021**, *15* (3), 2000385.

(68) *Lumerical Inc.,* https://www.lumerical.com/products/; access date **2020**. (accessed.

(69) Paniagua-Domínguez, R.; Yu, Y. F.; Miroshnichenko, A. E.; Krivitsky, L. A.; Fu, Y. H.; Valuckas, V.; Gonzaga, L.; Toh, Y. T.; Kay, A. Y. S.; Luk'yanchuk, B. Generalized Brewster effect in dielectric metasurfaces. *Nature communications* **2016**, *7* (1), 1-9.